\documentclass[a4paper]{article}

\usepackage[utf8]{inputenc}

\usepackage{multicol}
\usepackage{amsthm}
\usepackage{bm}

\usepackage[dvipdfmx]{graphicx}

\usepackage{bmpsize}
\usepackage{amssymb}
\usepackage{amsmath}
\usepackage{color}
\usepackage{amsthm}
\usepackage{subcaption} 

\newcommand{\BA}{\begin{eqnarray}}
\newcommand{\EA}{\end{eqnarray}}

\definecolor{dgreen}{rgb}{0.0, 0.5, 0.0}

\begin{document}

\fontsize{14pt}{16.5pt}\selectfont

\begin{center}
\bf{A Generalization for Ultradiscrete Limit Cycles\\ in a Certain Type of Max-Plus Dynamical Systems
}
\end{center}
\fontsize{12pt}{11pt}\selectfont
\begin{center}
Shousuke Ohmori$^{1,2*)}$ and Yoshihiro Yamazaki$^{3)}$\\ 
\end{center}

\bigskip

\noindent
\it{
1)~National Institute of Technology, Gunma College, 
    580 Toribamachi, Maebashi-shi, Gunma 371-8530, Japan.
}\\
\it{
2)~Waseda Research Institute for Science and Engineering, Waseda University, 
    Shinjuku, Tokyo 169-8555, Japan.
}\\
\it{
3)~Department of Physics, Waseda University, 
    Shinjuku, Tokyo 169-8555, Japan.
}

\bigskip

\noindent
*corresponding author: 42261timemachine@ruri.waseda.jp\\
~~\\
\rm
\fontsize{11pt}{14pt}\selectfont\noindent

\baselineskip 30pt

{\bf Abstract}\\
%
Dynamical properties of a generalized max-plus model 
for ultradiscrete limit cycles are investigated. 
This model includes both the negative feedback model 
and the Sel'kov model.
It exhibits the Neimark-Sacker bifurcation, 
and possesses stable and unstable ultradiscrete limit cycles.
The number of discrete states in the limit cycles 
can be analytically determined 
and its approximate relation is proposed.
Additionally, relationship between the max-plus model 
and the two-dimensional normal form 
of the border collision bifurcation is discussed.




%
%
\section{Introduction}
\label{sec:intro}

Difference equations utilizing max-plus algebra are powerful tools 
for describing nonlinear and nonequilibrium phenomena. 
These descriptions have been applied in various systems: 
soliton behavior in integrable systems\cite{Tokihiro2004}, 
traffic flow in social systems\cite{Nishinari1998}, 
inflammatory response in physiological systems\cite{Carstea2006}, 
reaction-diffusion dynamics in dissipative systems
\cite{Murata2013,Matsuya2015,Ohmori2016}, 
feedback mechanisms in biochemical systems\cite{Gibo2015}, 
and bifurcation phenomena in dynamical systems\cite{Ohmori2020}. 
A max-plus difference equation can be systematically constructed 
using the ultradiscretization method\cite{Tokihiro2004}, 
which has been successfully applied in integrable systems. 
Furthermore, this method has proven effective in non-integrable systems. 
The cases of the max-plus negative feedback model 
and the max-plus Sel'kov model are typical examples\cite{Ohmori2021, Yamazaki2021, Ohmori2022, Ohmori2023}.
The crucial point is that the dynamical properties 
of the original (continuous and discrete) systems can be retained 
in the max-plus models, 
which are derived from the original ones via ultradiscretization.
%
%
%
Actually, the discrete limit cycles 
in the tropically discretized models can be retained 
in their ultradiscretized max-plus ones\cite{Yamazaki2023}. 
In this manuscript, we report dynamical properties 
for the following set of max-plus difference equations, 
\begin{eqnarray}
    \left\{ \,
    \begin{aligned}
     X_{n+1} & = & Y_n+\max (0, T X_n), \\
     Y_{n+1} & = & B-\max (0, D X_n),
    \end{aligned}
    \right.
    \label{eqn:mp_model}
\end{eqnarray}
where $T \geq 0$, $D \geq 0$, 
and $B$ takes an arbitrary real value.
It is noted that eq.(\ref{eqn:mp_model}) includes 
both the max-plus negative feedback model 
and the max-plus Sel'kov model.
In the next section, 
we review dynamical properties of the above two models 
as the special cases of eq.(\ref{eqn:mp_model}).
In sec.\ref{sec.2},  we report the properties 
for the solution flow of eq.(\ref{eqn:mp_model}) 
focusing on its piecewise linearity.
In sec.\ref{sec.3}, we discuss the relation 
between the numbers of the states in the ultradiscrete limit cycle 
and the parameters $(T, D)$ in eq.(\ref{eqn:mp_model}).
The discussion and conclusion are given 
in Secs. \ref{sec.4} and \ref{sec.6}, respectively.

\section{Review for the special cases of eq.(\ref{eqn:mp_model})}

Here we review the previous results for 
the following two special cases of eq.(\ref{eqn:mp_model}).

\subsection*{case I: $T=0$ and $D>0$}

In this case, eq.(\ref{eqn:mp_model}) corresponds 
to the max-plus negative feedback model\cite{Ohmori2023}.
(i) For $B<0$, eq.(\ref{eqn:mp_model}) has 
a single stable fixed point. 
Note that any initial state converges to the fixed point 
at most four iteration steps for $D>0$.
(ii) When $B>0$, the fixed point of eq.(\ref{eqn:mp_model}) 
behaves as the spiral sink for $0<D<1$, 
and the spiral source for $D > 1$. 
Additionally in the case of $D > 1$, 
there exist the stable and the unstable limit cycles 
which consist of four discrete states.
Figure \ref{fig:lc_NFandSelkov} (a) shows 
the stable (blue) and the unstable (red) limit cycles 
with $B = 1$, $D = 1.5$, and $T=0$.
Note that the number of the discrete states 
composing of the limit cycles is independent of the value of $D$.

\subsection*{case II: $T=D$}

\begin{figure}[t!]
    \begin{center}
        \includegraphics[width=5cm]{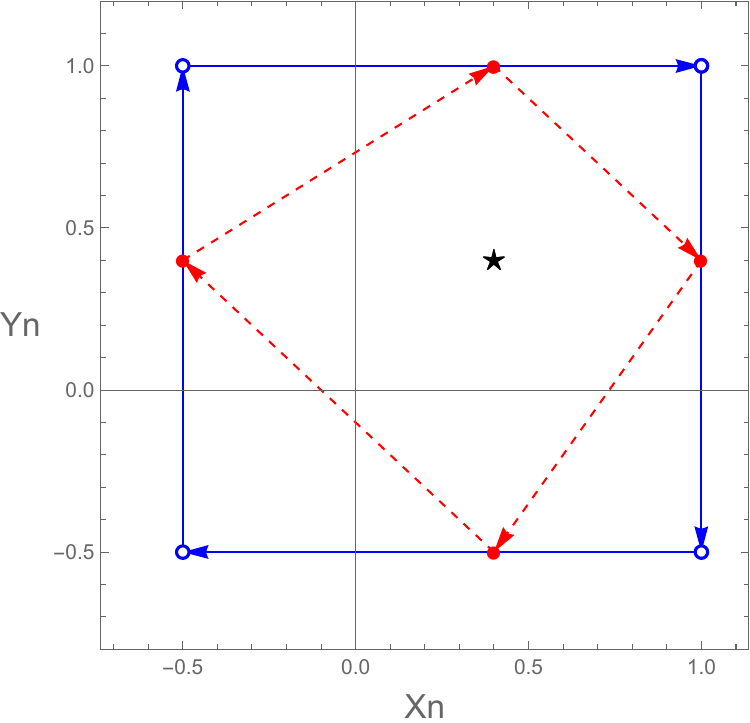}
        \hspace{2cm}
        \includegraphics[width=5cm]{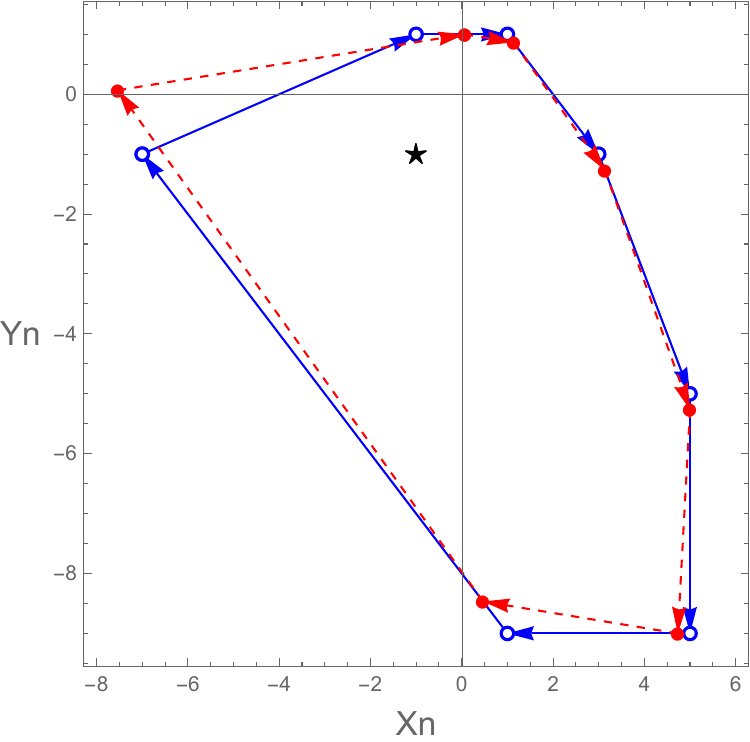}
        \\
        \hspace{0cm}
        (a)
        \hspace{6cm}
        (b)\\
        \caption{\label{fig:lc_NFandSelkov} 
            The two limit cycles $\mathcal{C}$ (blue circles) 
            and $\mathcal{C}_s$ (red circles) found 
            in the max-plus equations for
            (a) the negative feedback model ($B=1, D=1.5, T=0$)\cite{Ohmori2023} and 
            (b) the Sel'kov model model ($B=1,D=T=2$)\cite{Ohmori2021}.
            Note that $\mathcal{C}$ is stable 
            and $\mathcal{C}_s$ is unstable.
            The black star in each figure shows the fixed point of each model. 
        }
    \end{center}
\end{figure}

When $T = D$, eq.(\ref{eqn:mp_model}) becomes the max-plus Sel'kov model\cite{Ohmori2021, Yamazaki2021}.
When $B>0$ and $T = D =2$, there exist the stable and the unstable limit cycles, 
which consist of seven different discrete states 
as shown in Fig.\ref{fig:lc_NFandSelkov} (b).
Furthermore for $T=D\equiv R$ ($R>1$),
we have obtained the relation between $R$ and the number 
of the discrete states in the limit cycles $p$\cite{Yamazaki2021}. 
Here we also introduce $n (= p -2)$; its meaning is explained 
at eq.(\ref{eqn:pn2}) in sec.\ref{sec.3}.
Actually, $R$-dependence of $p$ 
is shown in the Tbl. \ref{tbl:periodicity}.
$R_{\text{min}}(n)$ and $R_{\text{max}}(n)$ show the minimum and the maximum values 
of $R$ for existence of the limit cycles with period $p (=n+2)$, respectively.
From Tbl.\ref{tbl:periodicity}, it is found that 
$p$ increases with increase of $R$. 
\begin{table}[h!]
	\begin{center}
		\caption{Regions of $R$ for existence of limit cycles 
            with period $p = n+2$.}
   		\begin{tabular}{c|c|l}
            $n$ & $p (= n + 2)$ & $R_{\text{min}}(n) \leq R\leq R_{\text{max}}(n)$ \\
            \hline
             4 &  6 & 1\color{white}.000000$\cdots $\color{black} $ \sim$ 1.83928$\cdots$ \\
             5 &  7 & 1.93318 $\cdots \sim$ 2.59205 $\cdots$\\
             6 &  8 & 2.60229 $\cdots \sim$ 2.99375 $\cdots$ \\
             7 &  9 & 2.99585 $\cdots \sim$ 3.24522 $\cdots$ \\
             8 & 10 & 3.24576 $\cdots \sim$ 3.41367 $\cdots$ \\
             9 & 11 & 3.41383 $\cdots \sim$ 3.53191 $\cdots$ \\
            10 & 12 & 3.53196 $\cdots \sim$ 3.61797 $\cdots$ \\
            \vdots & \vdots & \hspace{1.9cm}\vdots
    \end{tabular}
		\label{tbl:periodicity}
	\end{center}
\end{table}
Besides, we have found that there exist a quasi-periodic structure 
when $R_{\text{max}}(n)< R < R_{\text{min}}(n+1)$\cite{Yamazaki2021}. 
Since $R_{\text{max}}(n)<  R_{\text{min}}(n+1)$ always holds for all $n$, 
the quasi-periodic structures can be found for all pairs $(n, n+1)$.
Note that such quasi-periodic structures do not exist 
in the max-plus negative feedback model ($T=0$, $D>0$). 

In both cases I and II, it is found that the occurrence of the pair of the stable and unstable limit cycles 
are due to phase lock caused by saddle-node bifurcation\cite{Yamazaki2023}. 
%

%
%

%
%

\section{Solution flow of eq.(\ref{eqn:mp_model})}
\label{sec.2}

To grasp the solution flow of eq.(\ref{eqn:mp_model}), 
let us divide the $X_n$-$Y_n$ plane into the two regions, 
(I) $X_n \geq 0$ and (II) $X_n < 0$.

\subsection{Piecewise linearity}
\label{sec.2.1}

First we consider the case where the state 
$\bm{x}_{n} = \left(
	\begin{array}{c}
		X_{n} \\
		Y_{n} 
	\end{array}
\right)$  
is in the region (II), $X_{n} < 0$.
Because of $T \geq 0$ and $D \geq 0$, 
the terms $\max(0, TX_{n})$ and $\max(0, DX_{n})$ 
in eq.(\ref{eqn:mp_model}) become zero.
Therefore, eq.(\ref{eqn:mp_model}) in the region (II) is represented as 
\begin{equation}
	\bm{x}_{n+1}
	=
	\left(
		\begin{array}{ccc}
			0 & 1  \\
			0 & 0  \\
		\end{array}
	\right)
	\bm{x}_{n}
	+
	\left(
		\begin{array}{ccc}
			0  \\
			B  
		\end{array}
	\right).  
	\label{eqn:ld_II}
\end{equation} 
Equation (\ref{eqn:ld_II}) is independent of $T$ and $D$, 
and has the unique fixed point $\bar{\bm{w}}_{2} \equiv (B,B)$. 
When the state $\bm{x}_{n}$ satisfies $X_{n} < 0$ and $Y_{n} \leq 0$, 
denoted as region (II)-1, the state reaches $\bar{\bm{w}}_{2}$ 
at two time steps: $(X_n, Y_n)\mapsto (Y_n, B) \mapsto (B, B)$.
When the state is in the region (II)-2, 
$X_{n} < 0$ and $Y_{n} > 0$, the state at the next step, $(Y_n, B)$, 
is in the region (I) since $Y_n>0$.

In the case where the state $\bm{x}_{n}$ is in the region (I), 
the state follows the following equation 
\begin{equation}
	\bm{x}_{n+1}
	= A \bm{x}_{n}
	+
	\left(
		\begin{array}{ccc}
			0  \\
			B  
		\end{array}
	\right), \;\;\;\text{where }
    A = \left(
		\begin{array}{ccc}
			T & 1  \\
			-D & 0  \\
		\end{array}
	\right).
	\label{eqn:ld_I}
\end{equation} 
The point $\bar{\bm{w}}_{1} \equiv \left(\frac{B}{1+D-T},\frac{(1-T)B}{1+D-T}\right)$ 
becomes the unique fixed point of eq.(\ref{eqn:ld_I}).
The dynamical properties of eq.(\ref{eqn:ld_I}) 
around $\bar{\bm{w}}_{1}$ are determined 
by $T = \mathrm{tr}A$ and $D = \mathrm{det}A$ as follows\cite{Galor}.
(i) For $D<T-1$, $\bar{\bm{w}}_{1}$ is saddle. 
(ii) For $T-1<D<\displaystyle \frac{T^2}{4}$, $\bar{\bm{w}}_{1}$ is node. 
(iii) For $D > \displaystyle \frac{T^2}{4}$, $\bar{\bm{w}}_{1}$ is spiral.
In the case of (iii), the stability of $\bar{\bm{w}}_{1}$ changes at $D=1$; 
it becomes stable for $D<1$ and unstable for $D>1$.
To investigate the dynamical properties of the limit cycles, 
we focus on $D>\displaystyle \frac{T^2}{4}$ and $D > 1$, 
where $D-T+1 > 0$ always holds.

\subsection{Trajectories}
\label{sec.2.2}

For eq.(\ref{eqn:mp_model}), 
the transformations $X_n/|B|\to X_n$ and $Y_n/|B|\to Y_n$ 
can be performed without essential change of its dynamical properties.
In other words, only the sign of $B$ is important 
for characterizing the dynamical behaviors of $\bm{x}_{n}$.
Therefore, the cases $B=\pm 1$ are considered hereafter.

\begin{figure}[t!]
    \begin{center}
        \includegraphics[width=5cm]{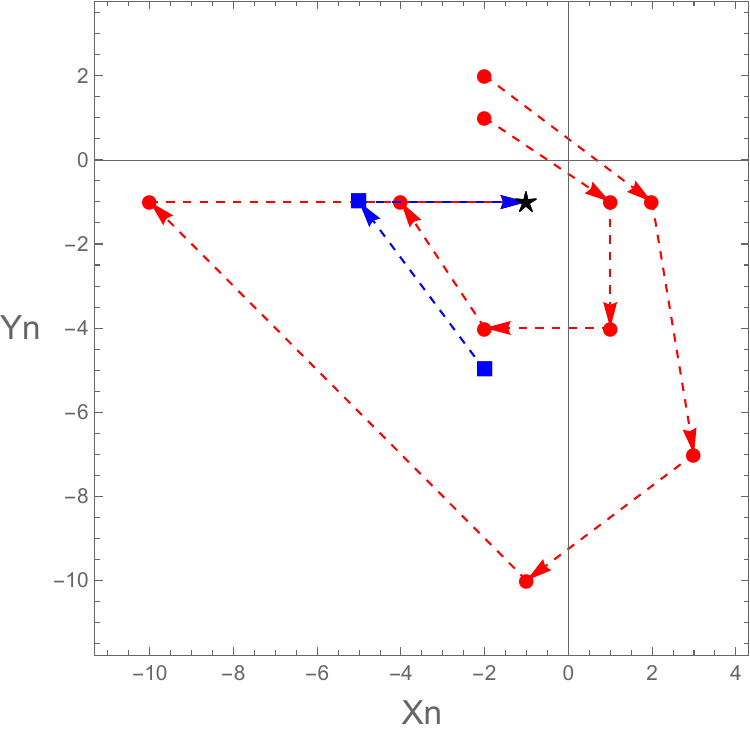}
        \hspace{2cm}
        \includegraphics[width=5cm]{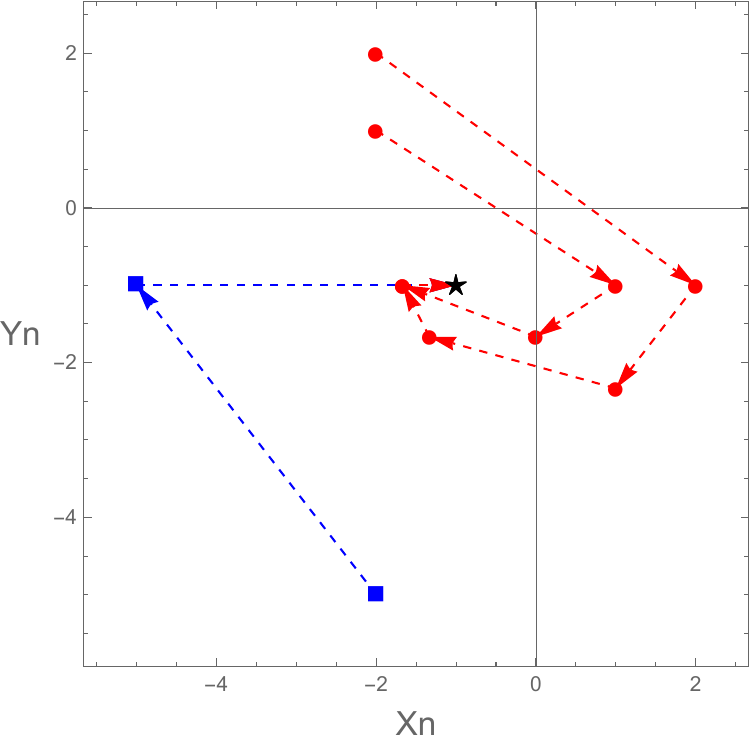}
        \\
        \hspace{0cm}
        (a)
        \hspace{6cm}
        (b)\\
        \caption{\label{fig:excitable} 
            Examples of trajectories obtained 
            from eq.(\ref{eqn:mp_model}) with $B=-1$. 
            (a) $(T,D)=(2,3)$, (b) $(T,D)=(1,2/3)$.
            One with the blue squares starts at a state in (II)-1, 
            and the others with the red circles start 
            at two different states in (II)-2.
            They reach the fixed point $(-1,-1)$ (the black star) 
            at the finite time steps. 
            Note that the red trajectories show excitability.
        }
    \end{center}
\end{figure}

When $B=-1$, both $\bar{\bm{w}}_{2}=(-1,-1)$ 
and $\bar{\bm{w}}_{1}=\left(-\frac{1}{1+D-T},-\frac{1-T}{1+D-T}\right)$ are in (II), 
then $\bar{\bm{w}}_{2}$ is considered as the fixed point 
of eq.(\ref{eqn:mp_model}).
As shown in sec. \ref{sec.2.1}, 
a state $(X_n,Y_n)$ in (II)-1 
reaches $\bar{\bm{w}}_{2}$ at the two steps. 
$(X_n,Y_n)$ in (II)-2 
gives $(X_{n+1},Y_{n+1})=(Y_n,-1)$ in (I).
Because of $D > \displaystyle \frac{T^2}{4}$, 
any state in (I) moves to rotate clockwise 
around $\bar{\bm{w}}_{1}$ and 
reaches a state in (II)-1 after finite time steps.  
%
Actually from eq.(\ref{eqn:ld_I}), $(X_n,Y_n)$ in (I) moves 
to $(X_{n+1},Y_{n+1}) = (TX_n + Y_n, -DX_n-1)$ at the next step, 
and $Y_{n+1} = -DX_n-1 < 0$ always holds since $X_{n} \geq 0$.
Thus, $\bar{\bm{w}}_{2}$ becomes a spiral sink for $B=-1$, 
and an excitablity can be confirmed with the initial state in (II)-2.
Figure \ref{fig:excitable} shows several trajectories 
obtained from eq.(\ref{eqn:mp_model}) with $B=-1$: 
(a) $(T,D)=(2,3)$ and (b) $(T,D)=(1,2/3)$.

When $B=+1$, $\bar{\bm{w}}_{2} = (1,1)$ 
and $\bar{\bm{w}}_{1}=\left(\frac{1}{1+D-T},\frac{1-T}{1+D-T}\right)$ are in (I), 
and $\bar{\bm{w}}_{1}$ becomes the fixed point of eq.(\ref{eqn:mp_model}).
For the dynamical properties in (II), 
$(X_n,Y_n)$ in (II)-1 reaches $\bar{\bm{w}}_{2}$ at two steps,  
and $(X_n,Y_n)$ in (II)-2 moves to a point $(Y_n, 1)$ 
on the line $\{(X,1); X > 0\}$ at the next step.
%
%
If $D<1$, $\bar{\bm{w}}_{1}$ becomes the stable spiral.
On the other hand if $D > 1$, 
$\bar{\bm{w}}_{1}$ becomes the unstable spiral, 
and when the state enters (II)-1 due to a clockwise rotational motion in (I), 
the state is reset to $\bar{\bm{w}}_{2}$.
Actually, the limit cycles  shown in Fig. \ref{fig:lc_NFandSelkov} 
are typical examples.
Thus, such a reset event plays the key role 
in generating the limit cycles for eq.(\ref{eqn:mp_model}).
%
%
Note that the stable limit cycle has the state $\bar{\bm{w}}_{2}$, 
whereas the unstable limit cycle does not have $\bar{\bm{w}}_{2}$.
Therefore the unstable limit cycle is unaffected 
by the above reset events.
%
%
%
In the following sections, we focus on these ultradiscrete limit cycles.  
%

\section{Number of states in the limit cycles}
\label{sec.3}

Now we fix $B=+1$.
%
%
Table \ref{tbl:p_numerical} shows the numerical esimations of $p$
as functions of (a) $T$ ($D=2$) and (b) $D$ ($T=2$).
For (a), from $D=2$, $T$ must satisfy $0\leq T < 2\sqrt{2}$.
For (b), from $T=2$, $D > 1$ must be satisfied.
Note that $T=D=2$ gives $p=7$ 
(the case of the max-plus Sel'kov model), 
and $T=0$, $D=2$ brings about $p=4$ 
(the case of the max-plus negative feedback model). 
\begin{table}[ht]
    \centering
    \caption{Numerical estimation of $p$, the number of states in a limit cycle, 
      as functions of (a) $T$ ($D=2$) and (b) $D$ ($T=2$).}
    \label{tbl:p_numerical}
    \begin{subtable}{.3\linewidth}
        \centering
        \caption{$D = 2$}
        \begin{tabular}{c|c}
            \hline
            $T$ & $p$ \\
            \hline
            0 & 4 \\
            1 & 5 \\
            2 & 7 \\
            2.5 & 10 \\
            2.8 & 26 \\
            \vdots & \vdots \\
            \hline
        \end{tabular}
    \end{subtable}%
    \begin{subtable}{.3\linewidth}
        \centering
        \caption{$T = 2$}
        \begin{tabular}{c|c}
            \hline
            $D$ & $p$ \\
            \hline
            1.1 & 16 \\
            1.5 &  8 \\
            2   &  7 \\
            4   &  5 \\
            10  &  4 \\
            50  &  4 \\
            \vdots & \vdots \\
            \hline
        \end{tabular}
    \end{subtable}
\end{table}
%
%
%
Based on the tendencies of $p$ against $D$ and $T$ shown in the Tbl.\ref{tbl:p_numerical}, 
the following properties can be elucidated.
(i) $p\geq 4$ holds for any $T \geq 0$ and $D > $.
Especially, $p = 4$ for large $D$ ($T \ne 0$) as well as $T=0$.
(ii) $p$ becomes larger (or goes to infinity) 
as $(T,D)$ approaches the relation $D=\frac{T^2}{4}$.
%

\begin{figure}[t!]
    \begin{center}
        \includegraphics[width=9cm]{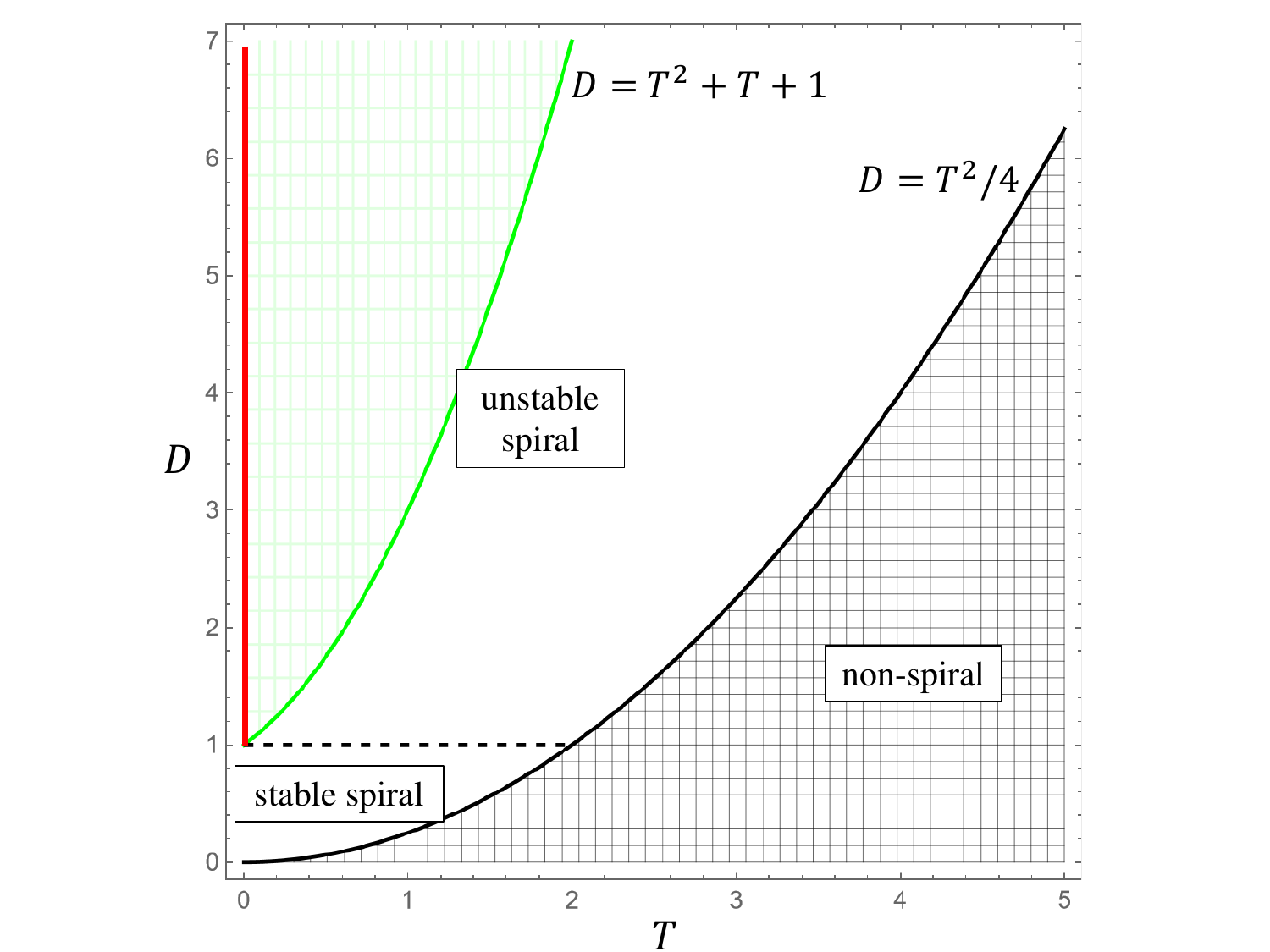}
        \caption{\label{fig:anti(i)} 
          Green area ($D>T^2+T+1$) shows the region for $p=4$. 
          The red line ($D>1$, $T=0$) corresponds to the case 
          of the max-plus negative feedback model.
        }
    \end{center}
\end{figure}

%
The property (i) can be confirmed as follows. 
Substituting $B=+1$ into eq.(\ref{eqn:ld_I}), 
eq.(\ref{eqn:mp_model}) in the region (I) 
is expressed as 
\begin{equation}
	\bm{x}_{n+1}
	=
	\left(
		\begin{array}{ccc}
			T & 1  \\
			-D & 0  \\
		\end{array}
	\right)
	\bm{x}_{n}
	+
	\left(
		\begin{array}{ccc}
			0  \\
			1  
		\end{array}
	\right), 
	\label{eqn:ld_I_B=1}
\end{equation} 
where the fixed point is $\bar{\bm{w}}_{1}=(\frac{1}{1+D-T},\frac{1-T}{1+D-T})$.
From eq.(\ref{eqn:ld_I_B=1}) with the initial state $\bm{x}_0=(1,1)$, 
we obtain $\bm{x}_1=(T+1,1-D)$ and $\bm{x}_2=(T^2+T+1-D,1-D(T+1))$.
Since $T>0$, $\bm{x}_1$ is in (I) for any $T$ and $D$.
For $\bm{x}_2$, when the condition 
\begin{eqnarray}
 T^2+T+1<D
 \label{condition1}
\end{eqnarray}
is satisfied, $\bm{x}_2$ belongs to (II)-1, i.e., $X_2<0$ and $Y_2<0$.
As shown above, every point in (II)-1 becomes $(1,1)=\bm{x}_0$ at two steps.
Then, the inequality (\ref{condition1}) shows the condition 
under which the limit cycle is composed of the following four discrete states: 
\begin{eqnarray}
    (1,1),(T+1,1-D),(T^2+T+1-D,1-D(T+1)),(1-D(T+1),1).
    \label{LCp4}
\end{eqnarray}
Figure \ref{fig:anti(i)} shows the $T$-$D$ graph for $p$, 
where the area satisfying the condition (\ref{condition1}) is painted with green. 
It is noted that the case of the max-plus negative feedback model ($T=0$, $D>1$) 
is included in this green region, which is shown as the red line in Fig. \ref{fig:anti(i)}.
%
%
%

As for the property (ii), we focus on the solution of eq.(\ref{eqn:ld_I_B=1}) 
after $n$ steps from the initial state $\bm{x}_0 =(1,1)$.
The solution $\bm{x}_n=(X_n(T,D), Y_n(T,D))$ can be explicitly written as 
\begin{eqnarray}
    X_n(T,D) & = & \frac{1}{(1+D-T)}+\frac{1}{2^{n+1}(1+D-T)i\sqrt{L}}
    \Biggl[(T^2-D(T+2))\biggl\{(T-i\sqrt{L})^n-(T+i\sqrt{L})^n\biggl\} 
    \nonumber\\
    & &
    +(D-T)\biggl\{(T-i\sqrt{L})^n
    +(T+i\sqrt{L})^n\biggl\}i\sqrt{L}
    \Biggl],
    \label{eqn:solution1}
\end{eqnarray}
\begin{eqnarray}
    Y_n(T,D) & = & \frac{1-T}{(1+D-T)}+\frac{1}{2^{n+1}(1+D-T)i\sqrt{L}}
    \Biggl[(D(2D-T))\biggl\{(T-i\sqrt{L})^n-(T+i\sqrt{L})^n\biggl\} 
    \nonumber\\
    & &
    + D\biggl\{(T-i\sqrt{L})^n
    +(T+i\sqrt{L})^n\biggl\}i\sqrt{L}
    \Biggl],
    \label{eqn:solution2}
\end{eqnarray}
where $L \equiv 4D-T^2 (>0)$.
For eqs.(\ref{eqn:solution1}) and (\ref{eqn:solution2}), 
the relation $Y_{n+1}(T,D)+DX_{n}(T,D)=1$ holds.
The state $\bm{x}_n$ also becomes the solution 
of eq.(\ref{eqn:mp_model}) when $X_{j} > 0$ holds 
for all $0 \leq j \leq n-1$.
In the case where the state first enters the region (II)-1 at the $n$-th step,
$X_n(T,D) < 0$ and $Y_n(T,D)\leq 0$, 
the state goes back to the initial state $\bm{x}_0$ after two steps 
as shown in sec.\ref{sec.2.1}.
Therefore, this trajectory becomes a periodic circuit 
and its period $p$ is given as 
\begin{equation}
  p=n+2.
  \label{eqn:pn2}
\end{equation}

\begin{figure}[t!]
    \begin{center}
        \includegraphics[width=7cm]{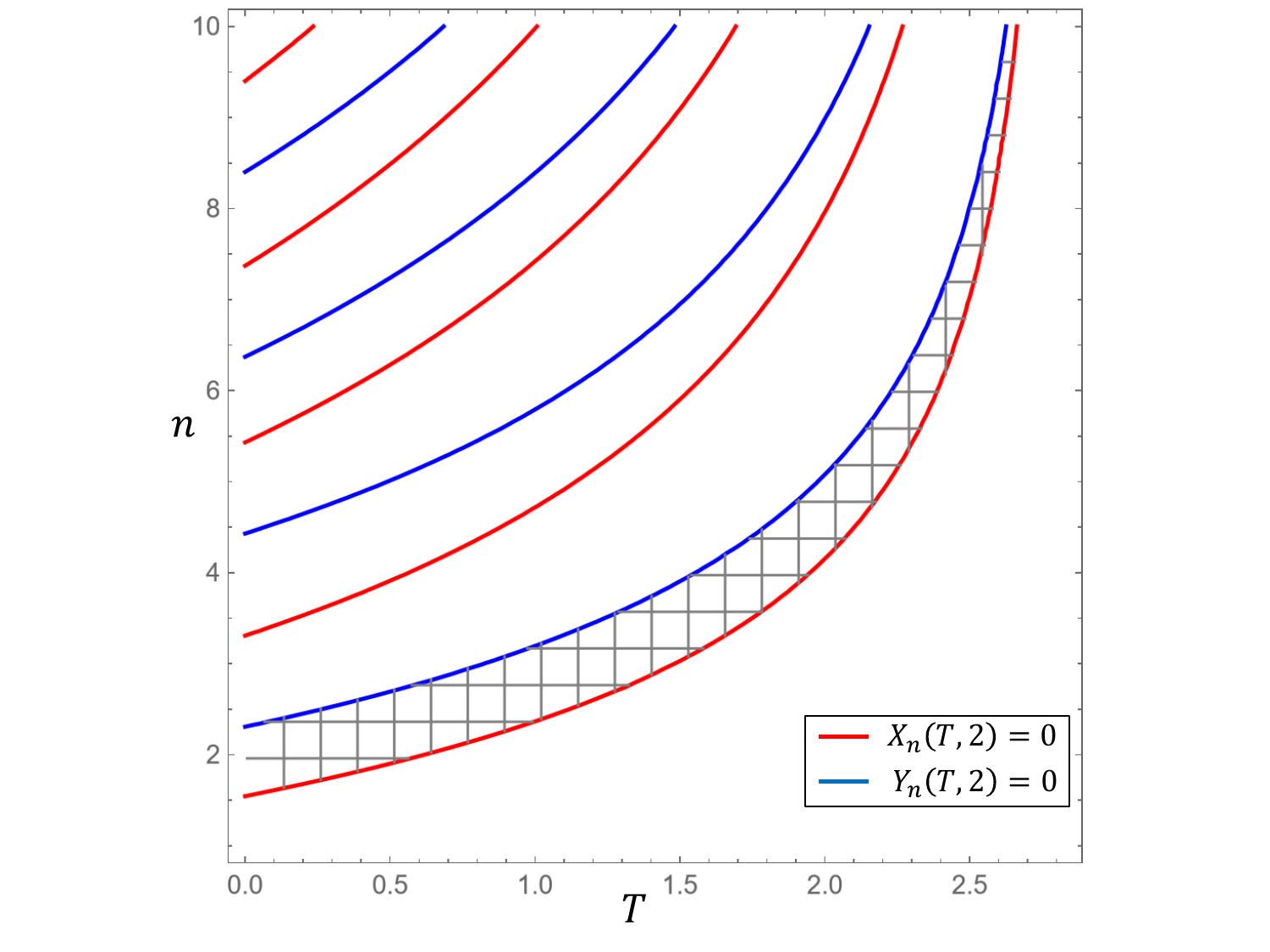}
        \hspace{1mm}
        \includegraphics[width=7cm]{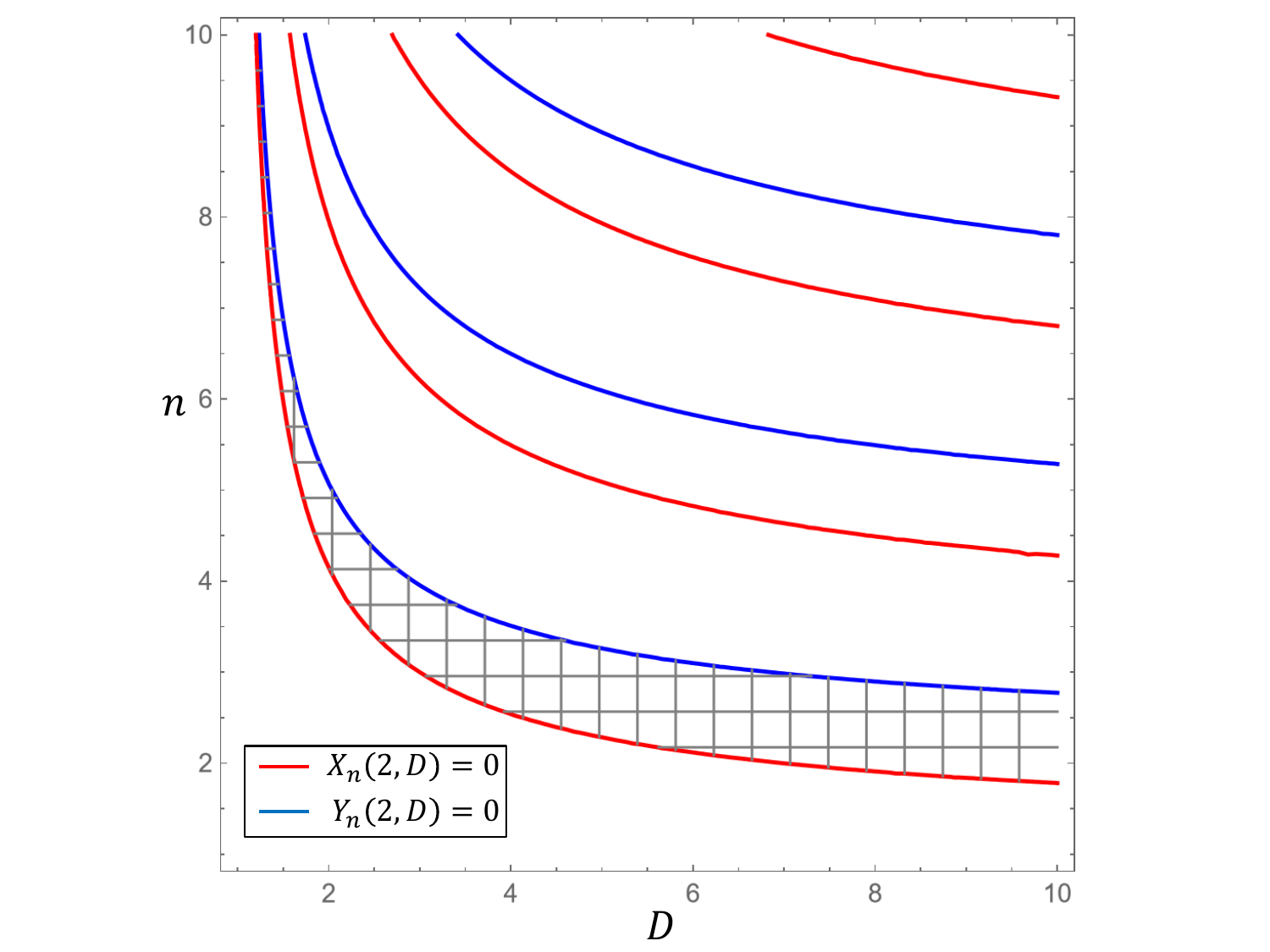}\\
        (a)
        \hspace{50mm}
        (b)
        \caption{\label{fig:anticipation(ii)_TD} 
        The conditions for the existence of limit cycles 
        when (a) $D=2$ and (b) $T=2$.
        These conditions depicted as the gray mesh regions  
        are enclosed by (a) $X_n(T,D=2)=0$ (red curve) and $Y_n(T,D=2)=0$ (blue curve)
        and by (b) $X_n(T=2,D)=0$ (red curve) and $Y_n(T=2,D)=0$ (blue curve). 
        }
    \end{center}
\end{figure}

For example, let us consider the case of $D=2$ (then $0\leq T < 2\sqrt{2}$).
Figure \ref{fig:anticipation(ii)_TD} (a) shows the contour plots 
of $X_n(T,D=2)=0$ (red) and $Y_n(T,D=2)=0$ (blue) as functions of $T$ and $n$.
It is found that the limit cycle can occur for $(n,T)$ 
in the gray mesh region of Fig. \ref{fig:anticipation(ii)_TD} (a). 
Denoting the solutions of $X_n(T,D=2)= 0$ and $Y_n(T,D=2)= 0$ 
with respect to $T$ by $T_X(n)$ and $T_Y(n)$,
the region of $T$ for occurrence of the limit cycle 
with period $p = n+2$ is 
shown as $T_Y(n)<T<T_X(n)$.
Actually $(n,T)=(2,0)$ and $(n,T)=(5,2)$ correspond to the cases of 
the max-plus negative feedback and 
the max-plus Sel'kov models, respectively.
Table \ref{tbl:D_fix_analysis} shows the numerical results 
of $T_X(n)$ and $T_Y(n)$, 
which are consistent with the results shown in Tbl.\ref{tbl:p_numerical}(a).
\begin{table}[h!]
	\begin{center}
		\caption{Numerically obtained $T_X(n)$ and $T_Y(n)$  
			for existence of limit cycles with period $p$.}
   		\begin{tabular}{c|c|l}
	    $n$ & $p(=n+2)$ &  $T_Y(n) \leq T\leq T_X(n)$ \\
			\hline
  	    2 & 4 & 0\color{white}.000000$\cdots $\color{black} $ \sim$ 0.61803$\cdots$ \\
  	    3 & 5 & 0.82287 $\cdots \sim$ 1.48119 $\cdots$\\
  	    4 & 6 & 1.55322 $\cdots \sim$ 1.94938 $\cdots$ \\
  	    5 & 7 & 1.97548 $\cdots \sim$ 2.22078 $\cdots$ \\
  	    6 &8  & 2.23088 $\cdots \sim$ 2.38848 $\cdots$ \\
  	    7 & 9 & 2.39266 $\cdots \sim$  2.49785 $\cdots$ \\
  	   8 & 10 & 2.49968 $\cdots \sim$ 2.57242 $\cdots$\\
  	   9 & 11 & 2.57326 $\cdots \sim$ 2.62517 $\cdots$ \\
  	 \vdots & \vdots & \hspace{1.75cm} \vdots
    \end{tabular}
		\label{tbl:D_fix_analysis}
	\end{center}
\end{table}

Next we consider the case of $T=2$ for another example.
Figure \ref{fig:anticipation(ii)_TD} (b) shows the contour plots 
of $X_n(T=2,D)=0$ (red) and $Y_n(T=2,D)=0$ (blue) 
as functions of $D$ and $n$.
The limit cycle can also occur for $(n,T)$ 
in the gray mesh region of Fig. \ref{fig:anticipation(ii)_TD} (b). 
In this case, a limit cycle with period $p=n+2$ occurs 
for $D_X(n)<D<D_Y(n)$, where $D_X(n)$ and $D_Y(n)$ are 
the solutions of $X_n(T=2,D)= 0$ and $Y_n(T=2,D)= 0$ with respect to $D$.
Table \ref{tbl:T_fix_analysis} shows the numerical results 
of $D_X(n)$ and $D_Y(n)$, 
which are consistent with the results shown in Tbl.\ref{tbl:p_numerical}(b).
Note that the point $(n,D)=(5,2)$ corresponds 
to the max-plus Sel'kov model.
\begin{table}[h!]
	\begin{center}
		\caption{Numerically obtained $D_X(n)$ and $D_Y(n)$
			for existence of limit cycles with period $p$.}
            \begin{tabular}{c|c|l}
	    $n$ & $p(=n+2)$ &  $D_X(n) \leq D\leq D_Y(n)$ \\
			\hline
  	    2 & 4 & 7.00000 $\cdots  \sim$ $+\infty \color{white}.000000\color{black}$ \\
  	    3 & 5 & 3.00000 $\cdots \sim$ 6.85410 $\cdots$\\
  	    4 & 6 & 2.07738 $\cdots \sim$ 2.93178 $\cdots$ \\
  	    5 & 7 & 1.69722 $\cdots \sim$ 2.03932 $\cdots$ \\
  	    6 & 8 & 1.49708 $\cdots \sim$ 1.67370 $\cdots$ \\
  	    7 & 9 & 1.37631 $\cdots \sim$  1.48149 $\cdots$ \\
  	   8 & 10 & 1.29682 $\cdots \sim$ 1.36545 $\cdots$\\
  	   9 & 11 & 1.24122 $\cdots \sim$ 1.28894 $\cdots$ \\
         \vdots & \vdots & \hspace{1.75cm} \vdots
    \end{tabular}
		\label{tbl:T_fix_analysis}
	\end{center}
\end{table}

\begin{figure}[t!]
    \begin{center}
        \includegraphics[width=7.5cm]{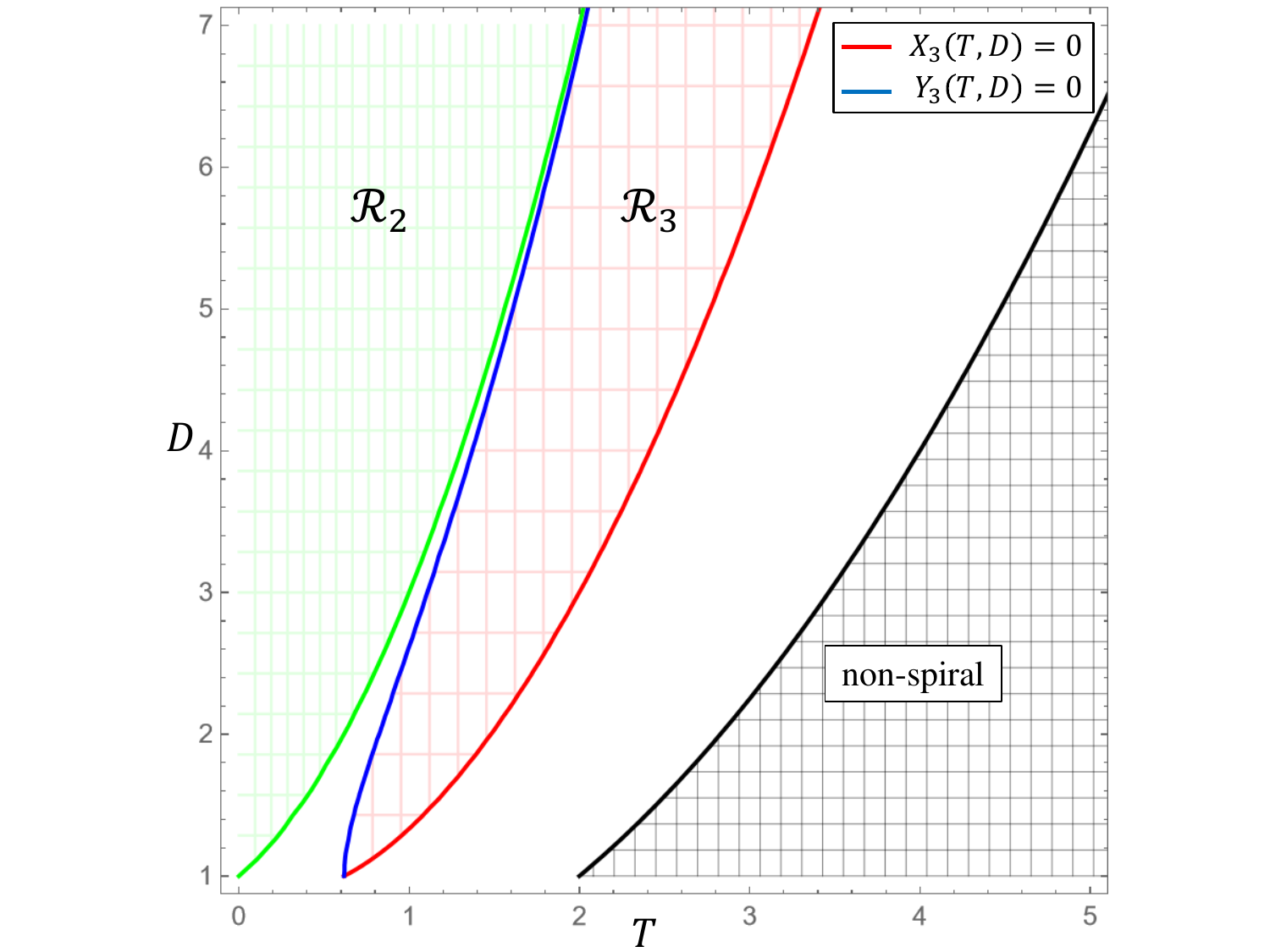}
        \hspace{1mm}
        \includegraphics[width=7.5cm]{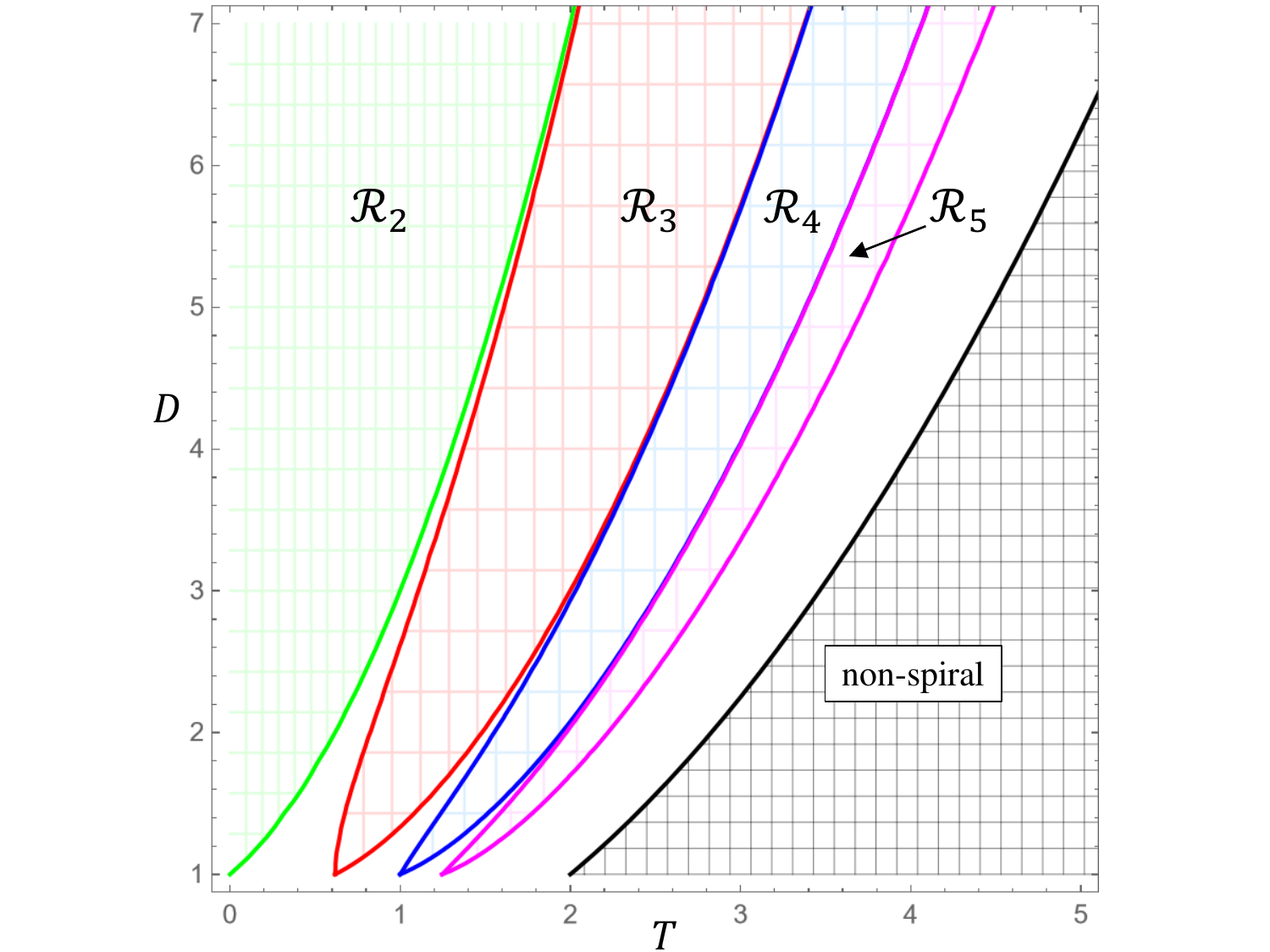}\\
        (a)
        \hspace{70mm}
        (b)
        \caption{\label{fig:lc_TD_domain} 
        (a) The regions $\mathcal{R}_2$ and $\mathcal{R}_3$.
        $\mathcal{R}_2$ is obtained as $(T, D)$ satisfying the condition (\ref{condition1}).
        $\mathcal{R}_3$ is enclosed by $X_3(T,D)=0$ (red) and $Y_3(T,D)=0$ (blue).
        (b) The regions $\mathcal{R}_2, \ldots, \mathcal{R}_5$.
        }
    \end{center}
\end{figure}

For each $n$, we can identify the region for the occurrence 
of the $p (=n+2)$ period limit cycle in the $(T,D)$-plane.
Such a region, denoted by $\mathcal{R}_n$, can be obtained 
as an area enclosed by the solution curves $X_n(T,D)= 0$ and $Y_n(T,D)= 0$.
Obviously $\mathcal{R}_2$ expresses the region 
satisfying eq. (\ref{condition1}), 
which is shown as the green region in Fig. \ref{fig:anti(i)}.
%
%
Figure \ref{fig:lc_TD_domain} (a) also shows the region $\mathcal{R}_3$ (red area), 
which satisfies $X_3(T,D) < 0$ and $Y_3(T,D)\leq 0$. 
%
%
Similarly, Fig. \ref{fig:lc_TD_domain} (b) shows 
the regions of $\mathcal{R}_2, \ldots, \mathcal{R}_5$, 
where the limit cycles with $2+2=4, \ldots, 5+2=7$ periods 
are obtained, respectively. 
Note that $\mathcal{R}_n$ exists for any integer $n \geq 2$.

For $\mathcal{R}_{n}$, we find the following properties.
(1) $\mathcal{R}_{n+1}$ emerges adjacent to the right of $\mathcal{R}_n$ 
as shown in Fig. \ref{fig:lc_TD_domain} (b).
(2) $\mathcal{R}_{2},\mathcal{R}_{3}, \ldots, \mathcal{R}_{\infty}$ successively 
appear and approach the curve $D=\frac{T^2}{4}$.
(3) There exists a gap between $\mathcal{R}_n$ and $\mathcal{R}_{n+1}$ for any $n$.  
%
%
When $(T,D)$ takes the value in the gap 
between $\mathcal{R}_n$ and $\mathcal{R}_{n+1}$, 
eq. (\ref{eqn:mp_model}) can possess the quasi-periodic cycle 
composed of $n+(n+1)=2n+1$ discrete states.
%
%
Note that the result in Tbl.\ref{tbl:periodicity} can be reproduced 
when we consider $T=D (\equiv R)$.
%
%
%
%
%
%


\section{Discussion}
\label{sec.4}

\subsection{Existence of unstable limit cycles}
\label{sec.4.1}

To find the state of an unstable limit cycle for eq.(\ref{eqn:mp_model}), 
we focus on the solution $\bm{x}_n=(X_n(Z,T,D), Y_n(Z,T,D))$ of eq.(\ref{eqn:ld_I_B=1}) 
by setting the initial state $\bm{x}_0=(Z,1)$ $(Z>0)$.
From eq.(\ref{eqn:ld_I_B=1}), $Y_{n+1}(Z,T,D)=-DX_{n}(Z,T,D)+1$ is satisfied 
when $X_{n}>0$.
When $\bm{x}_{n+1}$ enters in (II), 
it is found from eq.(\ref{eqn:ld_II}) 
that $\bm{x}_{n+2}$ is given as $(Y_{n+1},1)$. 
Hence, the solution of the equation $Z = Y_{n+1}$, namely 
\begin{eqnarray}
    Z = -DX_{n}(Z,T,D)+1 
    \label{eqn:Zcondition}
\end{eqnarray}
with respect to $Z$ brings about a periodic trajectory; 
we denote the solution as $Z_n(T,D)$.
%
%
%
Based on the constraint $X_n(Z,T,D)>0$, 
$Z_n(T,D)$ must satisfy $0 < Z_n(T,D) < 1$.
%
%

\begin{figure}[t!]
    \begin{center}
        \includegraphics[width=7.5cm]{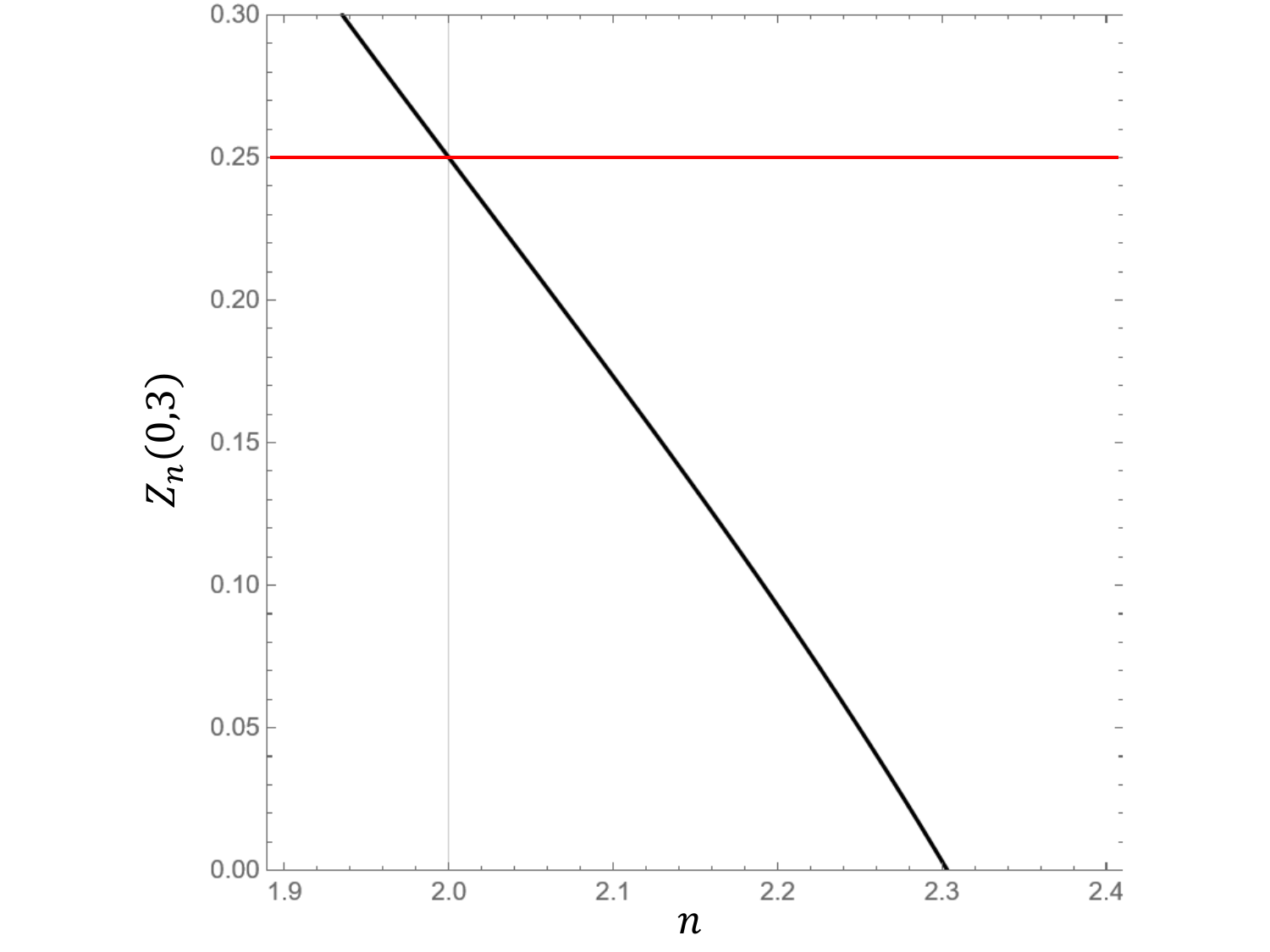}
        \hspace{1mm}
        \includegraphics[width=7.5cm]{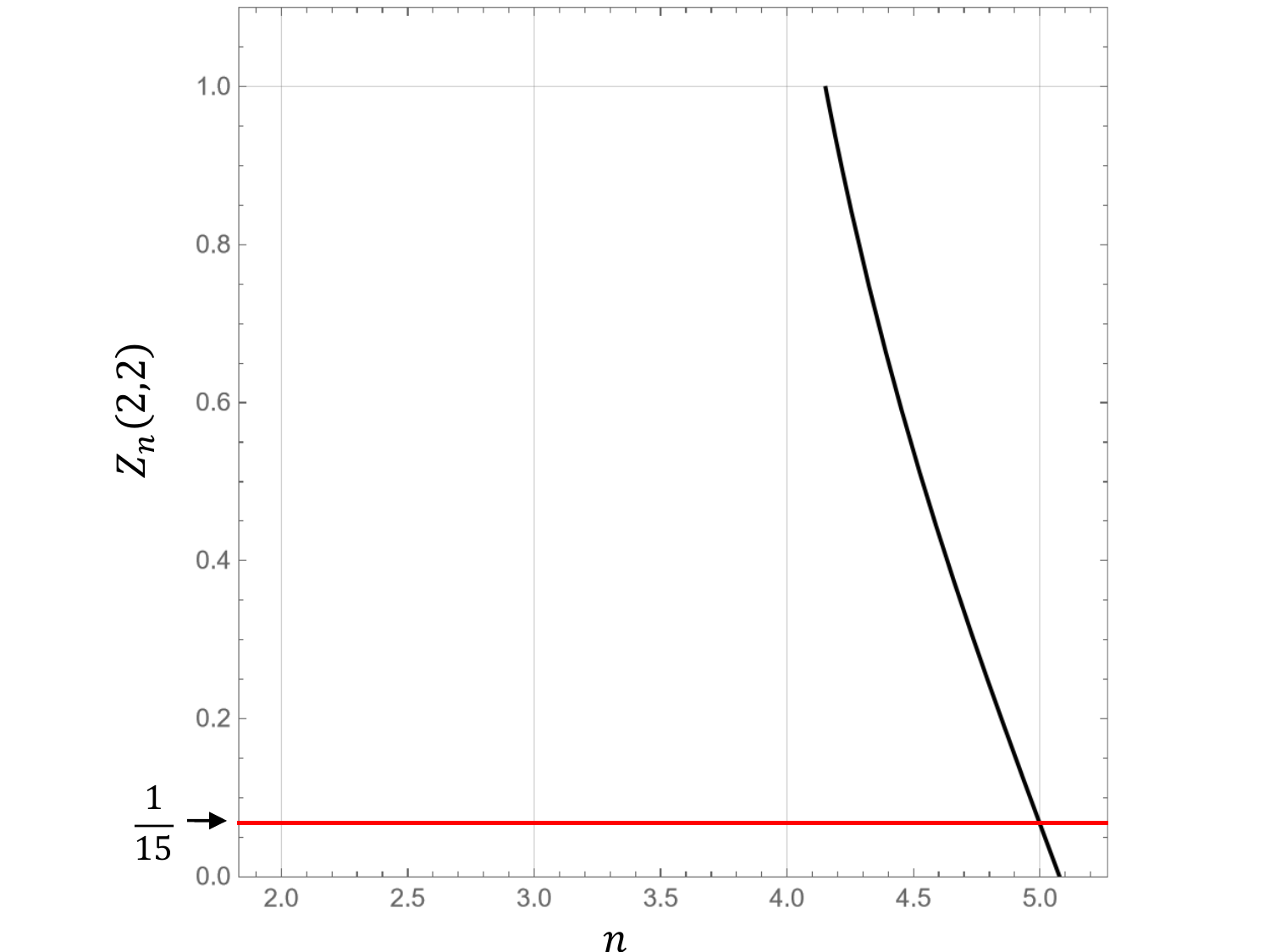}\\
        (a)
        \hspace{70mm}
        (b)
        \caption{\label{fig:uns.l.s.2-2} 
        The graphs of (a) $Z_n(T=0,D=3)$ for the negative feedback model and (b) $Z_n(T=2,D=2)$ for the Sel'kov model.
        }
    \end{center}
\end{figure}

Here we denote $Z_n(T,D)$ with the minimum time step $n$ as $Z_s$, 
which also satisfies $0 < Z_s < 1$. 
Note that $(Z_s, 1)$ is included in the ultradiscrete limit cycle.
Actually from the previous results\cite{Ohmori2023}, 
$Z_s(T=0,D>1) = \frac{1}{D+1}$ for the max-plus negative feedback model 
and $Z_s(T=2,D=2) = \frac{1}{15}$ for the max-plus Sel'kov model.
Figure \ref{fig:uns.l.s.2-2} shows the graphs of 
(a) $Z_n(T=0,D=3)$ and (b) $Z_n(T=2,D=2)$.
From the Fig. \ref{fig:uns.l.s.2-2} (a), it is found that 
the minimum step $n$ satisfying $0 < Z_n(T=0,D=3) < 1$ is 
$n=2$ ($p = n+2 =4$), 
and we obtain $Z_s=0.25= \frac{1}{D+1}$.
Then the periodic orbit with the state $\left(\frac{1}{D+1}, 1\right)$ 
is found to be the unstable limit cycle $\mathcal{C}_s$ shown in Fig. \ref{fig:lc_NFandSelkov} (a).  
Similarly, Fig. \ref{fig:uns.l.s.2-2} (b) shows 
that $n=5$ ($p = n+2 = 7$) is the minimum step satisfying $0 < Z_n(T=2,D=2) < 1$ 
and we obtain $Z_s=\frac{1}{15}$.
It is confirmed that the periodic orbit with the state 
$\left(\frac{1}{15}, 1\right)$ composing of $p=7$ states is identical 
to the unstable limit cycle $\mathcal{C}_s$ shown in Fig. \ref{fig:lc_NFandSelkov} (b).  
%
%
%

\begin{figure}[t!]
    \begin{center}
        \includegraphics[width=12cm]{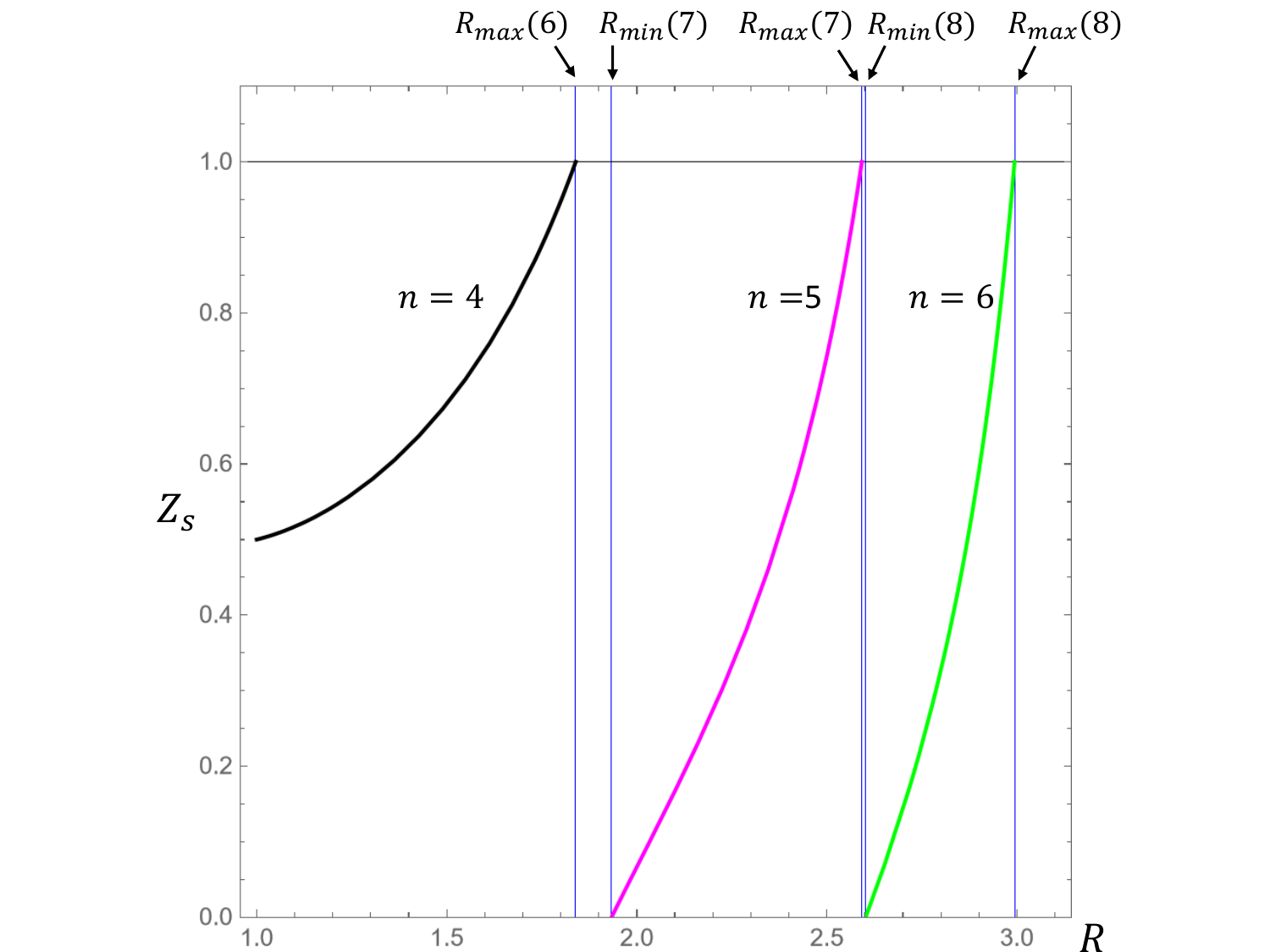}
        \caption{\label{fig:unstablelc_R} 
        $Z_s(R)$ for the generalized max-plus Sel'kov model 
        when $n=4, 5, 6$.
        }
    \end{center}
\end{figure}

We next consider the generalized max-plus Sel'kov model: $T=D=R$.
In this case, $Z_s(T=R,D=R)$ becomes a function of $R$. 
Figure \ref{fig:unstablelc_R} shows the graphs of $Z_s$ obtained numerically 
for $n=4$ (black), $n=5$ (magenta), and $n=6$ (green).
It is found that for each $n$ the value of $Z_s$ increases up to $Z_s=1$ as $R$ increases, 
suggesting that the unstable limit cycle approaches the stable limit cycle 
having the state $(1, 1)$ with increase of $R$. 
Note that for $n\geq 4$, the values of $R$ for $Z_s=0$ and $Z_s=1$ 
correspond to $R_{\text{min}}(n)$ and $R_{\text{max}}(n)$, 
respectively as shown in Tbl. \ref{tbl:periodicity}.
From Fig. \ref{fig:unstablelc_R}, it is also found that 
when $R_{\text{max}}(n)<R<R_{\text{min}}(n+1)$, 
the quasi periodic cycle appears 
instead of a pair of stabel and unstable limit cycles.
%


The stability of the ultradiscrete limit cycle 
containing $(Z_s,1)$ can be understood as follows.
Now we set the initial state $\bm{x}_0=(Z_s + \epsilon,1)$ 
on the line $\{(X,1);X>0\}$.
If $X_{1}, X_{2}, \ldots, X_{n} > 0$ and $\bm{x}_{n+1}$ exists 
in the region (II)-2, $X_{n+1} < 0$ and $Y_{n+1} > 0$, 
then the state $\bm{x}_{n+2} = (Y_{n+1}, 1)$ lies
on the line from eq.(2).

For eq.(\ref{eqn:ld_I_B=1}), 
\[
	\bm{x}_{n+1}
	= A \bm{x}_{n}
	+
	\left(
		\begin{array}{ccc}
			0  \\
			1  
		\end{array}
	\right), \;\;\;\text{where }
    A = \left(
		\begin{array}{ccc}
			T & 1  \\
			-D & 0  \\
		\end{array}
	\right), 
\]
if we divide the initial state $\bm{x}_0$ 
into the two parts, $\bm{x}^{s}_0 + (\epsilon, 0)$, 
where $\bm{x}^{s}_0 \equiv (Z_s, 1)$, 
the state after $n+1$ steps can be described as
\begin{equation}
	\bm{x}_{n+1} = \bm{x}^{s}_{n+1} 
      + \epsilon \bm{u}_{n+1}, \;\;\;\text{where }
    \bm{u}_{n+1} = A^{n+1} 
	\left(
		\begin{array}{ccc}
			1  \\
			0  
		\end{array}
	\right).
	\label{eqn:unst_1}
\end{equation} 
Here, considering 
$\bm{x}_n = (X_n, Y_n)$, $\bm{x}^{s}_n \equiv (X^s_n, Y^s_n)$, 
$\bm{u}_{n} \equiv (u_n, v_n)$, and $Y^{s}_{n+1}=Z_s$, 
we obtain 
\begin{equation}
	Y_{n+1} = Y^{s}_{n+1} + \epsilon v_{n}
	  = Z_{s} + \epsilon v_{n+1}.
	\label{eqn:unst_1_Y}
\end{equation} 
Similarly, if we divide $\bm{x}_0$ 
into $\bar{\bm{x}}_0+(Z_s + \epsilon, 0)$, 
where $\bar{\bm{x}}_0 \equiv (0, 1)$, 
the state after $n+1$ steps can be also described as
\begin{equation}
	\bm{x}_{n+1} = \bar{\bm{x}}_{n+1} 
      + (Z_s + \epsilon) \bm{u}_{n+1}.
	\label{eqn:unst_2}
\end{equation} 
Note that $\bar{\bm{x}}_{1} = (1, 1)$ and 
$\bar{\bm{x}}_{n+1}$ is found to be equal to 
$\left( X_n(T, D), Y_n(T, D) \right)$, which are given 
by eqs.(\ref{eqn:solution1}) and (\ref{eqn:solution2}).
For $Y_{n+1}$, from eq.(\ref{eqn:unst_2}), we obtain 
\begin{equation}
	Y_{n+1} = Y_n(T, D) + (Z_s + \epsilon) v_{n+1}.
	\label{eqn:unst_2_Y}
\end{equation} 
From eqs.(\ref{eqn:unst_1_Y}) and (\ref{eqn:unst_2_Y}), 
the following relation holds:
\begin{equation}
	Y_n(T, D) = (1-v_{n+1}) Z_{s}.
	\label{eqn:unst_3}
\end{equation} 
Since $Y_n(T, D) \leq 0$ and $Z_{s} > 0$, 
$v_{n+1}$ must be grater than or equal to 1.
$v_{n+1} = 1$ corresponds to the case 
where the limit cycle with the point $(Z_s, 1)$ is neutrally stable.
When $v_{n+1} >1$, a point in the vicinity of $(Z_s,1)$ moves away 
from $(Z_s,1)$, and the limit cycle is unstable.
%

\begin{figure}[t!]
    \begin{center}
        \includegraphics[width=9cm]{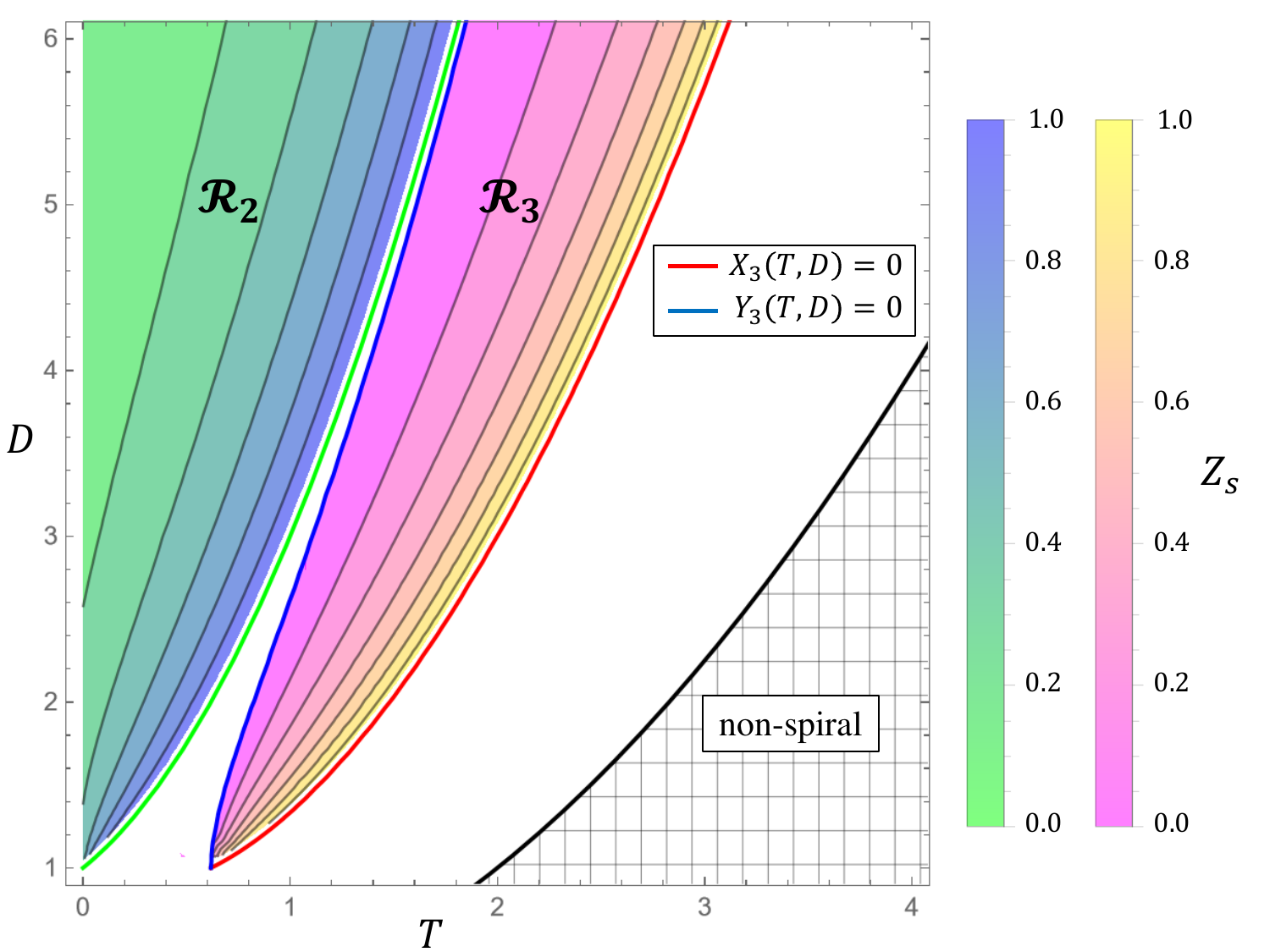}
        \caption{\label{fig:unstablelc_TD_domain} 
        Contour plots of $Z_n(T,D)=Z_s$ for $n=2$ and $n=3$, 
        which are superimposed on Fig. \ref{fig:lc_TD_domain} (a).
        The color map from green to blue in the region $\mathcal{R}_2$ 
        shows the value of $Z_s$ for $n=2$, and 
        the color map from pink to yellow in the region $\mathcal{R}_3$ 
        shows the value of $Z_s$ for $n=3$.
        }
    \end{center}
\end{figure}

Regarding the $(T,D)$-dependence of $Z_s$, 
Fig. \ref{fig:unstablelc_TD_domain} shows the numerical results 
for $n=2$ and $n=3$, which are drawn superimposed 
on Fig. \ref{fig:lc_TD_domain} (a). 
%
%
It is found that regions where the unstable limit cycles exist for $n=2$ and $n=3$ 
are completely identical to $\mathcal{R}_2$ and $\mathcal{R}_3$, respectively.
Thus, for eq. (\ref{eqn:mp_model}), limit cycles always appear in pairs, 
stable and unstable with the same period.
The stable and the unstable limit cycles in pairs 
tend to coalesce when $(T,D)$ gets close to $X_n(T,D)=0$,  
one of the boundaries of $\mathcal{R}_n$, 
since $Z_s$ approaches to $1$.
Moreover between $\mathcal{R}_n$ and $\mathcal{R}_{n+1}$, 
the pair of the limit cycles disappears and 
the quasi periodic orbit emerges.
Especially for the case of the negative feedback model and the Sel'kov model, 
we have already revealed that 
the emergence of the pair of stable and unstable limit cycles 
is inherited from their tropically discretized models, 
which possess ultradiscrete states due to phase lock caused by the saddle node bifurcation 
\cite{Yamazaki2023,Yamazaki2023b}.
Such inheritance of the dynamical properties is considered 
to be valid to the general case, eq.(\ref{eqn:mp_model}).

\subsection{Approximate estimation of $n$ as a function of $T$ and $D$}
\label{sec.4.2}

\begin{figure}[t!]
    \begin{center}
        \includegraphics[width=10cm]{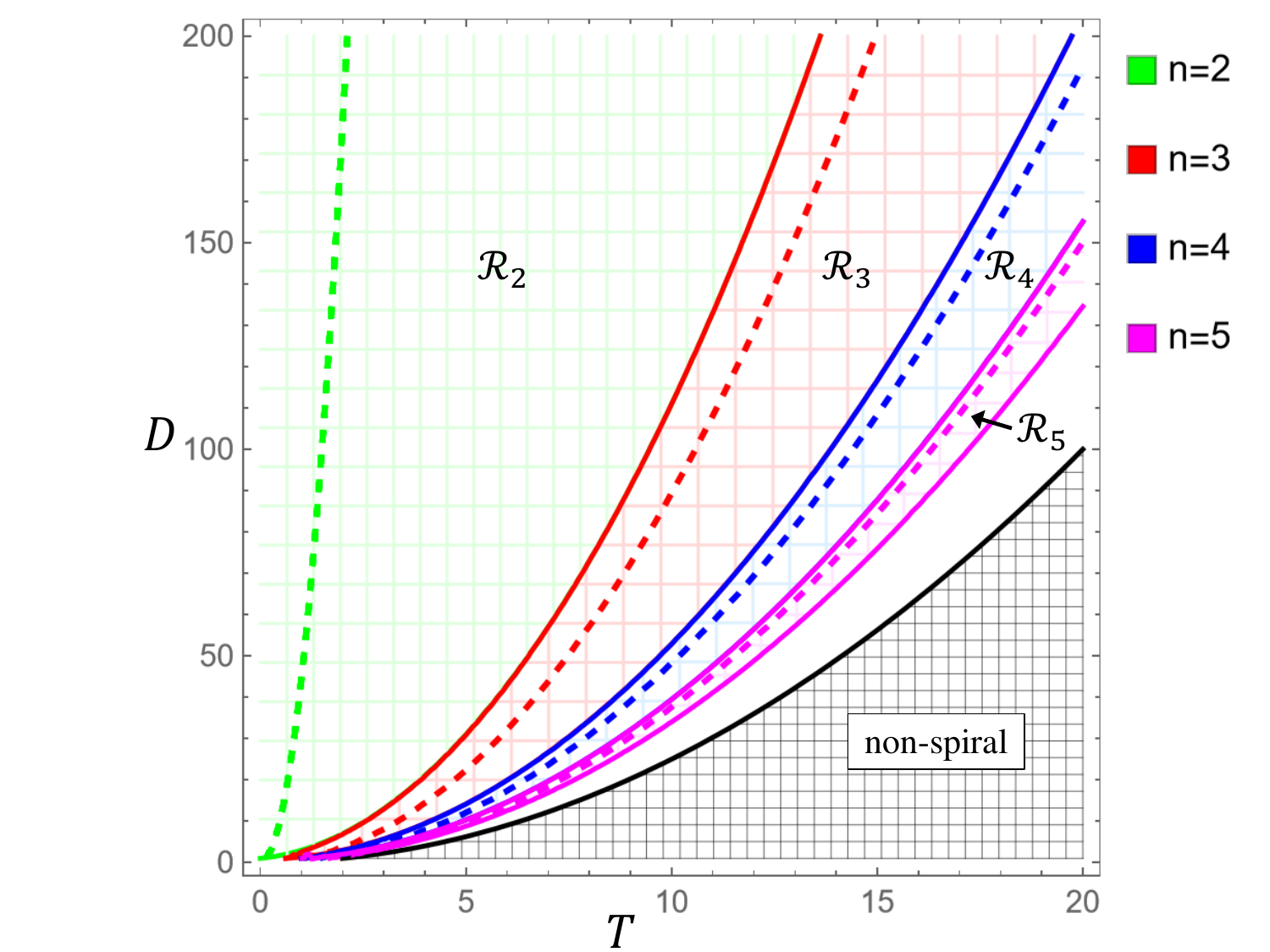}
        \caption{\label{fig:approx_TD} 
        The broken curves show the approximate relations given by eq.(\ref{eqn:approx_DT}) 
        for $n = 2,3,4,5$.
        They are plotted with $\mathcal{R}_2, \ldots, \mathcal{R}_5$ 
        shown in Fig. \ref{fig:lc_TD_domain} (b). 
        Here we set $c = -0.1$ in eq.(\ref{eqn:approx_DT}).
        }
    \end{center}
\end{figure}

For the max-plus Sel'kov model ($T=D$ in eq.(\ref{eqn:mp_model})), 
we have already obtained the approximate relation between $n$ and $T$ 
in the previous paper\cite{Yamazaki2021}: 
\begin{equation}
    \cos \frac{\pi}{n-c} = \frac{\sqrt{T}}{2}, 
    \label{eqn:approx_Selkov_nT}
\end{equation}
where $c$ is a constant. 
Now in a similar way to the previous approach, 
the following variable transformation 
from $(T, D)$ to $\theta$ is considered: 
\begin{equation}
    e^{i\theta} \equiv \frac{T + i\sqrt{4D-T^2}}{2\sqrt{D}}.
    \label{eqn:intro_theta}
\end{equation}
Applying this variable transformation 
to eqs.(\ref{eqn:solution1}) and (\ref{eqn:solution2}), 
it is found that these equations include 
the following terms: $\sin n\theta$ and $\sin (n-1)\theta$.
Then the relations $X_n(T, D) = 0$ and $Y_n(T, D)=0$ can be 
approximately satisfied when $\sin (n-c) \theta = 0$.
As the smallest positive $n$ for $\sin (n-c)\theta = 0$, 
$(n-c)\theta = \pi$ is obtained.
Therefore, the approximate relation between $n$ and $(T, D)$ 
is shown as 
\begin{equation}
    \cos \theta = \cos \frac{\pi}{n-c} = \frac{T}{2\sqrt{D}},  
    \label{eqn:approx_costheta}
\end{equation}
or 
\begin{equation}
    D = \frac{1}{4 \cos^{2} \frac{\pi}{n-c}} T^{2}.
    \label{eqn:approx_DT}
\end{equation}
Figure \ref{fig:approx_TD} shows the results for 
comparison of eq.(\ref{eqn:approx_DT}) with $\mathcal{R}_n$ 
for $n=2, \ldots, 5$.
It is found that each approximate relation is in each region.
%

\subsection{Relevance to border collision bifurcation}
\label{sec.4.3}

As stated in sec.\ref{sec.2.1}, 
eq.(\ref{eqn:mp_model}) is considered as the piecewise linear dynamical system 
with the bifurcation parameter $B$.
The interesting point is that the fixed point for eq.(\ref{eqn:mp_model}) 
changes by the sign of $B$.
Now we set $D > 1$. 
For $B<0$, the fixed point is $(B,B)$, 
and it behaves as the stable node.
On the other hand, for $B>0$, 
the fixed point switches to $(B,-B)$, which is the unstable spiral.
Therefore, eq.(\ref{eqn:mp_model}) brings 
about a bifurcation at $B=0$ by switching of the two different fixed points, 
and eq.(\ref{eqn:mp_model}) can possess limit cycles for $B>0$.
%
%
These dynamical properties are known 
as the border collision bifurcation (BCB)\cite{Bernardo2008}.

To discuss the relevance to BCB, 
we consider the following max-plus model, 
which corresponds to a generalization of eq.(\ref{eqn:mp_model}) : 
%
%
%
%
%
\begin{eqnarray}
    \left\{ \,
    \begin{aligned}
    X_{n+1} & =  Y_n+\max (T' X_n, T X_n), \\
    Y_{n+1} & =  B-\max (D' X_n, D X_n),
    \end{aligned}
    \right.
    \label{eqn:generalized_mp_model}
\end{eqnarray}
where we assume 
\begin{eqnarray}
    0\leq T'\leq T,~0\leq D'\leq D.
    \label{mp_condition}
\end{eqnarray}
When $T'=D'=0$, eq. (\ref{eqn:generalized_mp_model}) 
is identical to eq. (\ref{eqn:mp_model}).
This max-plus model can be obtained from the continuous dynamical system, 
$\displaystyle\frac{dx}{dt}=-x+(x^{T'}+x^{T})y$ and $\displaystyle\frac{dy}{dt}=b-(x^{D'}+x^{D})y$, 
via tropical discretization\cite{Murata2013} 
and ultradiscretization\cite{Tokihiro2004}. 
Now performing the transformation, 
\begin{eqnarray}
    X_n \to X_n, ~Y_n-B \to Y_n,
    \label{eqn:BCBtransformations}
\end{eqnarray}
eq.(\ref{eqn:generalized_mp_model}) becomes 
\begin{eqnarray}
    \left\{ \,
    \begin{aligned}
    X_{n+1} & = Y_n+\max (T' X_n, T X_n)+B, \\
    Y_{n+1} & = -\max (D' X_n, D X_n),
    \end{aligned}
    \right.
    \label{eqn:generalized_mp_model2}
\end{eqnarray}
which is exactly the same as the normal form of BCB in two dimensional case\cite{Banerjee1999},
\begin{eqnarray}
    &
    \left(
		\begin{array}{ccc}
			X_{n+1}  \\
			Y_{n+1}  
		\end{array}
	\right)
    =  
    \left\{ \,
    \begin{aligned}
    \left(
		\begin{array}{ccc}
			T' & 1  \\
			-D' & 0  \\
		\end{array}
	\right)
	\left(
		\begin{array}{ccc}
			X_n  \\
			Y_n  
		\end{array}
	\right)
	+
	\left(
		\begin{array}{ccc}
			B  \\
			0  
		\end{array}
	\right) ~~(X_n<0),
    \\
    \left(
		\begin{array}{ccc}
			T & 1  \\
			-D & 0  \\
		\end{array}
	\right)
	\left(
		\begin{array}{ccc}
			X_n  \\
			Y_n  
		\end{array}
	\right)
	+
	\left(
		\begin{array}{ccc}
			B  \\
			0  
		\end{array}
	\right) ~~(X_n>0).
    \end{aligned}
    \right.
    \label{eqn:BCBnormalform}
\end{eqnarray}
Since the transformation (\ref{eqn:BCBtransformations}) is 
the piecewise topologically conjugate, 
the dynamical properties of eq.(\ref{eqn:generalized_mp_model}) 
are completely consistent with those of eq.(\ref{eqn:BCBnormalform}).
Then, eq.(\ref{eqn:mp_model}) can be considered 
as the special case of eq.(\ref{eqn:BCBnormalform}).
In the report by Banerjee and Grebogi \cite{Banerjee1999}, 
the dynamical properties of eq.(\ref{eqn:BCBnormalform}) have been characterized 
by classifing the parameter space $(T', T, D', D)$ 
as shown in Fig.6 of ref.\cite{Banerjee1999}.
In this parameter space, the case of eq.(\ref{eqn:mp_model}) 
corresponds to the subspace $T'=D'=0$, $0\leq T$, and $0\leq D$, 
which is now denoted as $\Omega$ hereafter.
%
%
When $T<2\sqrt{D}$, $\Omega$ corresponds 
to the lower bounded line of {\it the spiral attractor to spiral attractor} region 
in their classification\cite{Banerjee1999}.
In this case, it is confirmed from Fig. \ref{fig:anti(i)} 
that eq. (\ref{eqn:mp_model}) possesses a spiral. 
When $2\sqrt{D}<T<1+D$, $\Omega$ becomes the lower bounded 
of {\it the regular attractor to regular attractor} region, 
in which eq. (\ref{eqn:mp_model}) has a node. 
When $T > 1+D$, $\Omega$ becomes lower bounded of 
{\it the Border collision pair bifurcation} region.
In this case, eq. (\ref{eqn:mp_model}) has a saddle.
%


Equations (\ref{eqn:mp_model}) and (\ref{eqn:generalized_mp_model})  
can provide an connection between BCB developed in the context of piecewise linear (smooth) dynamical systems 
and the ultradiscrete bifurcation developed 
in the field of tropically discretized dynamical systems 
and their ultradiscretized max-plus dynamical systems.
The emerging connection identified in our present discussion 
is expected to offer potential insights and advancements 
in both fields. 
%

\section{Conclusion}
\label{sec.6}

We have reported the dynamical properties 
of the max-plus discrete model, eq.(\ref{eqn:mp_model}), 
which is considered as a general model 
including both the negative feedback model and the Sel'kov model.
Focusing on its piecewise linearity, 
eq.(\ref{eqn:mp_model}) exhibits the Neimark-Sacker bifurcation 
and possesses ultradiscrete limit cycles.
The solution flow, or trajectory, of the limit cycles 
can be characterized by the two roles: rotation and reset. 
Note that the limit cycles emerge in pairs; stable and unstable.
We have also identified (i) the relation 
between the period $p$ of the limit cycles and the values of $T$ and $D$, 
(ii) the $(T, D)$ region for existence of the quasi periodic structures, 
and (iii) the approximate relation between $n$ (or $p - 2$) and $(T, D)$.
Furthermore, eq.(\ref{eqn:mp_model}) can be understood 
as the special case of the two dimensional normal form 
of the border collision bifurcation.
%

%

\bigskip

\noindent
{\bf Acknowledgement}

The authors are grateful to Prof. M. Murata, Prof. K. Matsuya, Prof. D. Takahashi, Prof. R. Willox, Prof. H.
Ujino, Prof. Y. Sato, Prof. A. Shudo, Prof. Emeritus Y. Aizawa, Prof. T. Yamamoto, and Prof. Emeritus A. Kitada for useful comments and encouragements.
This work was supported by JSPS
KAKENHI Grant Numbers 22K13963 and 22K03442.

\bigskip

\noindent
{\bf Data Availability}

Data sharing is not applicable to this article as no new data were created or analyzed in this study.

\end{document}